\documentclass[11pt,a4paper]{article}
\usepackage{jcappub}
\usepackage{xtab}

\newcommand{\be}{\begin{equation}}
\newcommand{\ee}{\end{equation}}
\newcommand{\rar}{\rightarrow}

\title{BBN with light dark matter}
\author{Zurab Berezhiani$^{a,b}$}
\author{Aleksander Dolgov$^{c,d,e,f}$}
\author{Igor Tkachev$^{c,g}$}

\affiliation{$^a$Dipartimento di Fisica, Universit\`a dell'Aquila, Via Vetoio, 67100 Coppito, L'Aquila, Italy} 
\affiliation{$^b$INFN, Laboratori Nazionali Gran Sasso, 67010 Assergi,  L'Aquila, Italy}
\affiliation{$^c$Laboratory of Cosmology and Elementary Particles, Novosibirsk State University, Pirogov street 2, 630090 Novosibirsk, Russia}
\affiliation{$^d$Dipartimento di Fisica, Universit`a degli Studi di Ferrara,  Polo Scientico e Tecnologico - Edicio C, Via Saragat 1, 44122 Ferrara, Italy}
\affiliation{$^e$Istituto Nazionale di Fisica Nucleare, Sezione di Ferrara,
Polo Scientiﬁco e Tecnologico - Ediﬁcio C, Via Saragat 1, 44122 Ferrara, Italy}
\affiliation{$^f$Institute of Theoretical and Experimental Physics,
Bolshaya Cheremushkinskaya ul. 25, 113259 Moscow, Russia}
\affiliation{$^g$Institute  for Nuclear Research of the Russian Academy of Sciences,
Moscow 117312, Russia}

\abstract{Effects of light millicharged dark matter particles on primordial nucleosynthesis are considered. It is shown that if the 
mass of such particles is much smaller than the electron mass, they lead to strong overproduction of Helium-4. An 
agreement with observations can be achieved by non-vanishing lepton asymmetry. Baryon-to-photon ratio at BBN 
and neutrino-to-photon ratio both at BBN and at recombination are noticeably different as compared to the standard 
cosmological model. The latter ratio and possible lepton asymmetry could be checked by Planck. For higher mass of new 
particles the effect is much less pronounced and may even have opposite sign.}

\begin{document}
\maketitle
\flushbottom

\section{Introduction}
\label{sec:intro}

During the  last year there arose  renewed interest to the impact of possible light dark matter particles on Big Bang Nucleosynthesis (BBN) 
and Cosmic Microwave Background radiation (CMB)~\cite{henning,boem,ho}. The interest was stimulated by an observation that the number 
of the effective neutrino species at BBN is possibly larger than that predicted by the standard model, $N_{eff} = 3.046$ (see e.g. Ref.~\cite{N-stand}). 
Indeed, according to the analysis of the Helium-4 abundance $N_{eff} = 3.7^{+0.8}_{-0.7} $~\cite{N-obs}. This unknown relativistic form of matter 
got  the name dark radiation.

The effect of additional particles on light element abundances is model dependent
and may lead both to a rise and to a decrease of $N_{eff} $. This has been already found in earlier works with MeV dark matter particles~\cite{kolb,Serpico:2004nm}. 
It was argued~\cite{Serpico:2004nm} that if the new particles, $X$, are in thermal equilibrium with neutrinos they  would lead to an increase of the Helium-4 fraction 
for $m_X \leq 10 $ MeV, while if they are in equilibrium with electrons, positrons and photons their effect is opposite. In the first case they correspond to an 
increase of $N_{eff}$ and in the second case $N_{eff}$ becomes lower.

In this paper we also consider an impact of new light particles on BBN. 
It is assumed that $X$-particles have a small electric charge, much smaller than the charge of electron, $e' \ll e$. 
We consider both cases of $m_X < m_e$ and $m_X> m_e$. 
The model with $m_X> m_e$ has been considered in detail previously~\cite{Serpico:2004nm}. In this paper we concentrate on the case $m_X < m_e$. As we shall see in what follows, during the BBN epoch these dark matter particles were in thermal equilibrium with  $e^+$, $ e^-$, and photons and decoupled from neutrinos.  Annihilation of $e^+e^-$ pairs proceeds both into photons and into dark matter pairs, $X\bar X$. Later on the bulk of dark matter particles annihilates into photons. Here we consider  the model where this happens after BBN,  but of course this should occur long before hydrogen recombination to avoid strong 
constraints on CMBR spectrum distortions. Light elements abundances in this model are  altered because of:
\begin{enumerate}
\item[i)] Increased Universe expansion rate during BBN;
\item[ii)] Increased neutrino temperature with respect to the photon temperature because the entropy released in $e^+e^-$-annihilation
is now distributed between photons and $X$-particles;
\item[iii)] Increased baryon-to-photon ratio during BBN.
\end{enumerate}
We will treat the influence of these factors on BBN analytically, and then calculate numerically in Sec.~\ref{sec:bbn}. But first let us define the presently allowed range of parameters of the model.

\section{Existing constraints on milli-charged particles}
\label{sec:constraints}

We start with the review of existing bounds on the mass and electromagnetic coupling, $\alpha'$, of milli-charged particles, in the range were they can be a (part of) dark matter.

\subsection{Laboratory bounds}

For $m_X < m_e$ the best particle physics bound on $\alpha'$ follows from the data on positronium decay to three photons~\cite{positronium}, according to which $e' < 3.4\cdot 10^{-5}\, e$, that is $\alpha_{11} \equiv 10^{11}(e')^2 /4\pi < 1$, see also ref.~\cite{Prinz}, where the similar limit was obtained. 
For very light particles, $m_X < 1~ {\rm keV}$, the best experimental limit $e'< 10^{-5}\, e$ comes  from the reactor experiments, \cite{Gninenko:2006fi}.
Heavier X-particles, $m_X > m_e$, are less restricted, e.g. for $m_X = 100$ MeV the bound is $\alpha' < 3\cdot 10^{-9}$~\cite{Prinz}. 
Stringent constraints on a MeV scale dark matter are provided by the fixed target experiments, in particular by the LSND \cite{Aguilar:2001ty}, see Ref.~\cite{Batell:2009di} for a detailed discussion. Corresponding bounds may not be  applicable to our model because
extra  light vector boson is not necessarily present and we do not aim to explain 0.511 
keV annihilation line. For the milli-charged scalar particles the main reaction to consider is $\pi^0 \rightarrow \gamma + X \bar X$. Branching of this process will be of order $\alpha'$, i.e. the number of produced $X$ particles will be $\alpha'$ times the number of produced neutrino. $X$ particles scatter in the detector with the cross-section $\sim \alpha \alpha' /E^2 $. It will be comparable to the weak cross-section at  $(E/ {\rm GeV})^4 \sim  10^{10} \alpha (\alpha')^2 \sim 10^{-14} \alpha_{11}^2$, i.e. at $E< {\rm MeV}\sqrt{\alpha_{11}}/3$, which looks safe.

\subsection{Constraints from stellar evolution}

The consideration of stellar cooling also allows to restrict the interaction strength of light $X$-particles~\cite{raffelt}. These particles could be abundantly produced in stellar cores and if their mean free path is shorter than the stellar radius, they would efficiently carry out stellar energy, strongly amplify the stellar luminosity, and  diminish the star age. Inside the Sun they could be created in the process $ ee \rar eeX \bar X$. The cross-section of this reaction can be roughly estimated as $\sigma (ee\rar eeX \bar X) \sim  \alpha^3 \alpha' /m_2^2$. The energy transferred to $X$-particles in the solar core per unit time is about 
\be 
L_{X}^{internal} \sim  \frac{4 \pi \alpha^3 \alpha'}{3 m_e^2}\, n_e^2 R^3_c E_X v_e \approx 
10^{44} \alpha_{11} \,{\rm erg/s}\,,
\label{L-X-int}
\ee 
where $n_e = 6\cdot 10^{25}/{\rm cm}^3$ is the density of electrons in the solar core, $R_c \approx 10^{10}$ cm
is the core radius, $E_X \sim $ keV is the $X$ energy, and $v_e \sim (T_c /m_e) \approx 0.05$ is the thermal velocity
of electrons in the solar core. However, the mean free path of $X$-particles is much shorter than $R_c$ and their
emission is suppressed by the factor $(l_{free}/R_c)^2$. The mean free path due to $eX$-scattering can be estimated as
\be
l_{free} = \left( \sigma_{eX} v_X n_e \right)\,,
\label{l-free}
\ee
where $\sigma_{eX} = 4\pi \alpha \alpha' L/ (m_X^2 v_X^4)$ is the Coulomb-like transport cross-section, and
$L$ is the Coulomb logarithm. Taking $m_X = 0.3$ keV and $v_X \sim 1$, we find $l_{free} \approx (4/L) $ cm.
Correspondingly the solar luminosity due to radiation of $X$-particles would be below $10^{-8} L_\odot$, where
$L_\odot = 4\cdot 10^{33}$ erg/sec is the solar luminosity.

The $X$-luminosity of white dwarfs (WD) can be estimated similarly. The electron density inside WD is $10^4$ larger than the solar one, while the radius is $10$ times smaller than $R_c$. So $L_{X}^{internal}$ would be $10^5$ times larger. However the mean free path of $X$-particles would be $10^3$ times smaller. The factor $10^{-4}$ comes from the larger $n_e$ and factor 10 comes from smaller $eX$-cross-section because of larger $X$ energy. So finally the X-particle  luminosity of white dwarfs would be about $10^{26}$ erg/sec, which looks safe.

There is a competing process of $X$-production for WD, namely decay of plasmons into  the $X\bar X$-pairs, because the plasma frequency in WD is about 10 keV and so 
it may be larger than $m_X$. However this process is less efficient than $X$-production in the above discussed reaction or at most comparable. Similar situation takes place in red giants.

Note that   surprisingly strong limit  $\alpha' < 10^{-30}$ for $m_X <10 $ keV   was obtained from the stellar cooling in ref.~\cite{davidson}.  However,  this bound is valid if $\alpha'< 10^{-18}$ is granted. The authors of ref.~\cite{davidson} assumed the validity of the last bound using BBN considerations. In our approach this BBN bound is invalid since we look if it is possible to allow millicharged dark matter particles at the expense of some  additional modification of BBN, e.g. by introduction of neutrino asymmetry. 

\subsection{Cosmological bounds}

Late annihilation of $X$-particles, when they go out of chemical equilibrium  with 
photons may distort the energy spectrum of CMB. The noticeable deviations from equilibrium occurred at $T \sim m_X/10$, when the bulk of $X$-particles was annihilating. At  this temperatures the energy density of $X$-particles is of the order of $4\cdot 10^{-2}\, \Omega_X / m_1$ of the energy density of the CMB photons. Kinetic equilibrium of energetic photons, created in $X \bar X$-annihilation, is restored by the elastic $\gamma\, e$-scattering very quickly, with the characteristic time of approximately $4\, ({\rm keV}/T)^3$ sec. However, chemical equilibration of photons demands inelastic reaction $\gamma\,e\rar 2\gamma\, e$ whose probability is approximately  five orders of magnitude smaller, so the effective reaction time is about $4\cdot 10^5$ sec  at $T= 1$ keV. It is quite close to the cosmological time at this temperature.
As is known~\cite{thermalization},  large non-equilibrium energy influx, of the same order of magnitude as the energy density of CMB, into the cosmological plasma would be perfectly thermalized if it took place at $z >10^7$. Smaller influx could be thermalized at lower $T$, so the chemical potential of the CMB  photons could be presumably reduced 
below the observational bound $\mu< 10^{-4}\, T$. 

The fraction of X-particles which were annihilating during recombination at $T \approx 0.2$~eV equals to  $0.2\; {\rm eV}/ (m_X/10)\sim 10^{-3} $ (for  S-wave and even smaller for P-wave annihilation). Therefore, the energy fraction of keV photons will be $10^{-3}$ times the ratio of energy densities of $X$-particles and photons during this epoch. If $\Omega_X$ is, say, 0.1 of the baryonic contribution to $\Omega$, then the corresponding perturbation of the chemical potential will not exceed $10^{-4}$, and therefore will not exceed observational bounds. Taking into account that spectrum distortions appear only after  photon degradation in energy, down from keV to the CMB energy, we see that the fraction of $X$-particles can be even higher.

Another very interesting and important bounds are coming from the analysis of CMBR anisotropies in the presence of millicharged particles \cite{milli-limits}.
One concludes that the fraction of such particles cannot be too large at the recombination if they are coupled to radiation. This bound is not really restrictive in  our situation since we are not assuming here that millicharged particles constitute significant fraction of matter presently.

\section{BBN in the presence of milli-charged particles}
\label{sec:bbn}

\subsection{Very light dark matter, $m_X < m_e$}

\subsubsection{Cosmological abundance}

Light charged particles will be in thermal equilibrium with electromagnetic plasma in the early Universe when 
\be
\sigma (e^+ e^- \rar X \bar X)\, n_e > H, 
\label{X-prod}
\ee
where 
\be
H =  \sqrt{\frac{8\pi^3 g_*}{90}}\, \frac{T^2 }{m_{Pl}^{~}} \approx \frac{5 T^2}{m_{Pl}^{~}}
\label{H}
\ee 
is the Hubble parameter at the radiation dominated cosmological stage and $n_e \sim T^3$ is the number density of electrons. Parameter $g_*= 10.75 + (7/4) (N_{eff}-3)$  counts the number of particle species at $ m_e < T < m_\mu$. The cross-section of $X\bar X$ production by electron-positron pairs for high electron energy, 
$E_e > m_e$, is
\be
\sigma (e^+ e^- \rar X \bar X) \sim \alpha \alpha' /T^2,
\label{sigma-ee-to-X}
\ee
where $\alpha' = (e')^2/4\pi$. So the process is in equilibrium roughly at $m_e < T < \alpha \alpha' m_{Pl}^{~}$ and we expect that at $T~\sim~1$ MeV $X$-particles have the equilibrium energy density, corresponding to one neutrino species up to the Bose-Fermi
factors. 

The late time (frozen) cosmological density of $X$-particles is determined by their annihilation into two photons and according to the standard calculations, see e.g. \cite{ad-zeld}, is equal to
\be
n_X^c \approx  \frac{10 n_\gamma  \ln [\sigma (X\bar X \rar 2\gamma) v_X m_X m_{Pl}]}
{\sigma (X\bar X\rar 2\gamma)  v_X m_X\, m_{Pl}}\, ,
\label{n-X}
\ee
where $v_X$ is the center-of-mass velocity of $X$-particles and 
\be
v \sigma (X \bar X \rar 2\gamma) = \pi (\alpha' )^2/ m_X^2\,.
\label{sigma-X-antiX-gamma}
\ee
So the mass density of $X$-particles at the present time, if they are stable, should be equal to
\be
\rho_X^c = 10 \,{\rm keV/cm}^3 \,\left(\frac{ m_1}{\alpha_{11}} \right)^2 \, [1 + 0.12\,\ln (\alpha_{11}^2/m_1)] , 
\label{rho-X-c}
\ee
where $m_1 = m_X/{\rm keV} $ and $\alpha_{11} = \alpha' /10^{-11}$. 
For comparison, the present day total cosmological energy density is about $5\, {\rm keV}/{\rm cm}^3$.

Estimating $\rho_X^c$ we have neglected the plasmon decay $\gamma_{pl} \rar X \bar X$. This is justified because at high temperatures, $T>m_e$,
when the plasma frequency, $\omega_{pl} \sim 0.1 T$, exceeds $m_X$ an account of the plasmon decay slightly
shifts equilibrium condition for $X$-particles which are in equilibrium anyhow. At smaller temperatures, $T<m_e$, when
$X\bar X$-annihilation is frozen, the plasma frequency is either Bolzmann suppressed, $\sim \exp (-m_e/m_X)$,
or suppressed by the smallness of $\alpha'$ and the plasmon decay may be neglected.

\subsubsection{Qualitative discussion of BBN}

Let us return now to evolution of $X$, $e^\pm$, and photons at BBN.
We assume that $X$-particles do not have anomalously strong interactions with neutrinos (though this possibility may be
interesting) and thus at $T$ below 1 MeV neutrinos are decoupled from $X$ and the electromagnetic part of the primeval
plasma. Due to reaction $e^+e^- \rar X \bar X$ and elastic scattering
$X$-particles remain in good thermal contact with electrons and positrons. 
Indeed, the ratio of the reaction rate to the Hubble parameter is
\be
\frac{1}{H}\,\frac{\dot n_X}{n_X} = \frac{\pi \alpha \alpha' n_e^2 \, m_{Pl}}{ 5 m_e^2  n_X T^2}
\label{dot-nX-nX}
\ee
and it is much larger than unity even for $T\ll m_e$, when the electron number density is exponentially suppressed. 
Correspondingly $e^+e^-$-annihilation equally well heats up photons and $X$-particles. So the entropy of electron-positron
pairs which in the usual case was totally transferred to photons, is now distributed between $\gamma$, $X$, and $\bar X$.
Correspondingly the entropy factor, which in the standard model was 11/4, now becomes 15/8 (assuming that $X$ have
spin zero). Hence the relative energy density of one neutrino species with respect to photons after $e^+e^-$-annihilation
instead of $(7/8)(4/11)^{4/3} = 0.227$ would become $(7/8)(8/15)^{4/3}=0.378 $. However, this change takes place asymptotically
at $T\ll m_e$ but at $T\sim m_e$ the effect is much smaller, which can be easily estimated analytically.

This rise of relative neutrino density has two-fold effect on BBN. First, rising density of $\nu_e$ would shift the temperature of
neutron-proton freezing to smaller values and thus leads to a decrease of $N_{eff}$. On the other hand, the rise of the energy density of
relativistic species increases $N_{eff}$ through the corresponding increase of the Hubble parameter at the same temperature. 
At the moment of $n/p$-freezing the first effect is stronger but later on the second effect 
dominates. It is especially pronounced in its impact on the time of the onset of BBN. The formation of first light elements started 
roughly at 
\be
T_{BBN} \approx E_b/\ln (1/\eta) \approx 70 \,{\rm keV},
\label{T-BBN}
\ee
where $\eta$ is the ratio of baryon to photon number densities, $\eta = n_B/n_\gamma$.
The time when $T_{BBN}$ is reached is determined by the number of relativistic species in the plasma according to 
$t_{BBN} \sim 1/(\sqrt{g_*} \,T)$. Now $t_{BBN}$ becomes considerably shorter and less neutrons have time to decay prior to BBN.
It can be described as quite large rise of $N_{eff}$. Below we present numerical calculations of primordial nuclei 
production for very light $X$-particles with $m_X < T_{BBN}$ which are in very good agreement with our simple analytical 
estimates.

The  baryon-to-photon ratio at BBN in this model is different from that determined from CMBR. The latter, according to
WMAP observations, is $\eta = 6.19 \times 10^{-10}$ \cite{Komatsu:2010fb}. We can determine $\eta_0$ at the moment of
$n/p$-freezing, which takes place at $T\approx 0.7 $ MeV using entropy conservation. Annihilation of 
$N$ extra degrees of freedom into photons during the epoch between BBN and recombination dilutes $\eta$  by the factor of $(2+N)/2$. 
Therefore, the  baryon-to-photon ratio right after $e^+e^-$ annihilation should be larger by this factor as compared to the standard cosmological 
model. Annihilation of $e^+e^-$ pairs leads to dilution factor of (11+2N)/(4+2N). The product of these two factors determines the initial condition for 
the baryon-to-photon ratio:
\be
\eta_0 = \frac{11+2N}{4}\; \eta.
\ee
In particular, with one charged extra scalar $N=2$ and $\eta_0 = (15/4)\, \eta$, while in the standard model this ratio is
 $\eta_0^{(st)} = (11/4)\, \eta$. One should keep in mind however, that in the standard model $\eta$ changes to its asymptotic value determined by CMB in the interval from $n/p$-freezing down to almost complete $e^+e^-$-annihilation, which takes place before $T=T_{BBN}$,  while in our scenario the ultimate entropy release from $X \bar X \rar 2\gamma $ takes place at $T\sim m_X \ll T_{BBN}$.

The ratio of the neutrino temperature to the photon one
at $n/p$-freezing in our model remains the same as in the standard case, $T_\nu/T_\gamma = (4/11)^{1/3}=0.71$,
while after $e^+e^-$-annihilation but prior to $T_{BBN}$ it rises as
\be
T_\nu = \left(\frac{4+2N}{11+2N}\right)^{1/3} \; T_\gamma,
\ee
so for scalar $X$-particles it would be $T_\nu/T_\gamma = (8/15)^{1/3} = 0.81$.
Subsequently after $X \bar X$-annihilation to photons this ratio drops down by 
$  [2/(2+N)]^{1/3} $ and turns to $T_\nu/ T_\gamma = [4/ (11+2N)]^{1/3} $ which for $N=2$
becomes $T_\nu / T_\gamma =  0.64$. In other words, the energy density of neutrinos determined by CMB would be 1.5 smaller than in the standard model.

\subsubsection{Numerical results}

\begin{figure}
\includegraphics[width=0.5\textwidth]{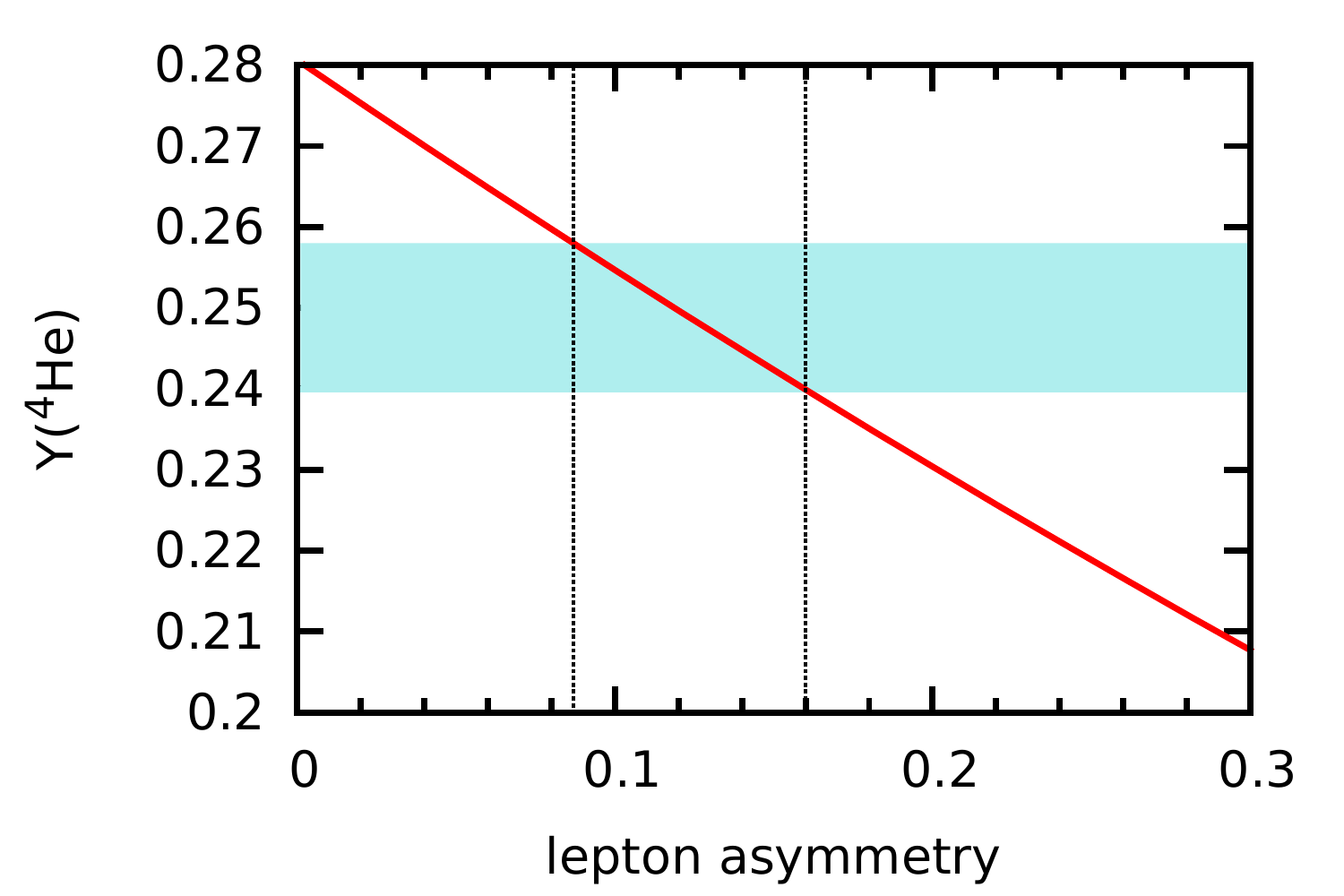}
\includegraphics[width=0.5\textwidth]{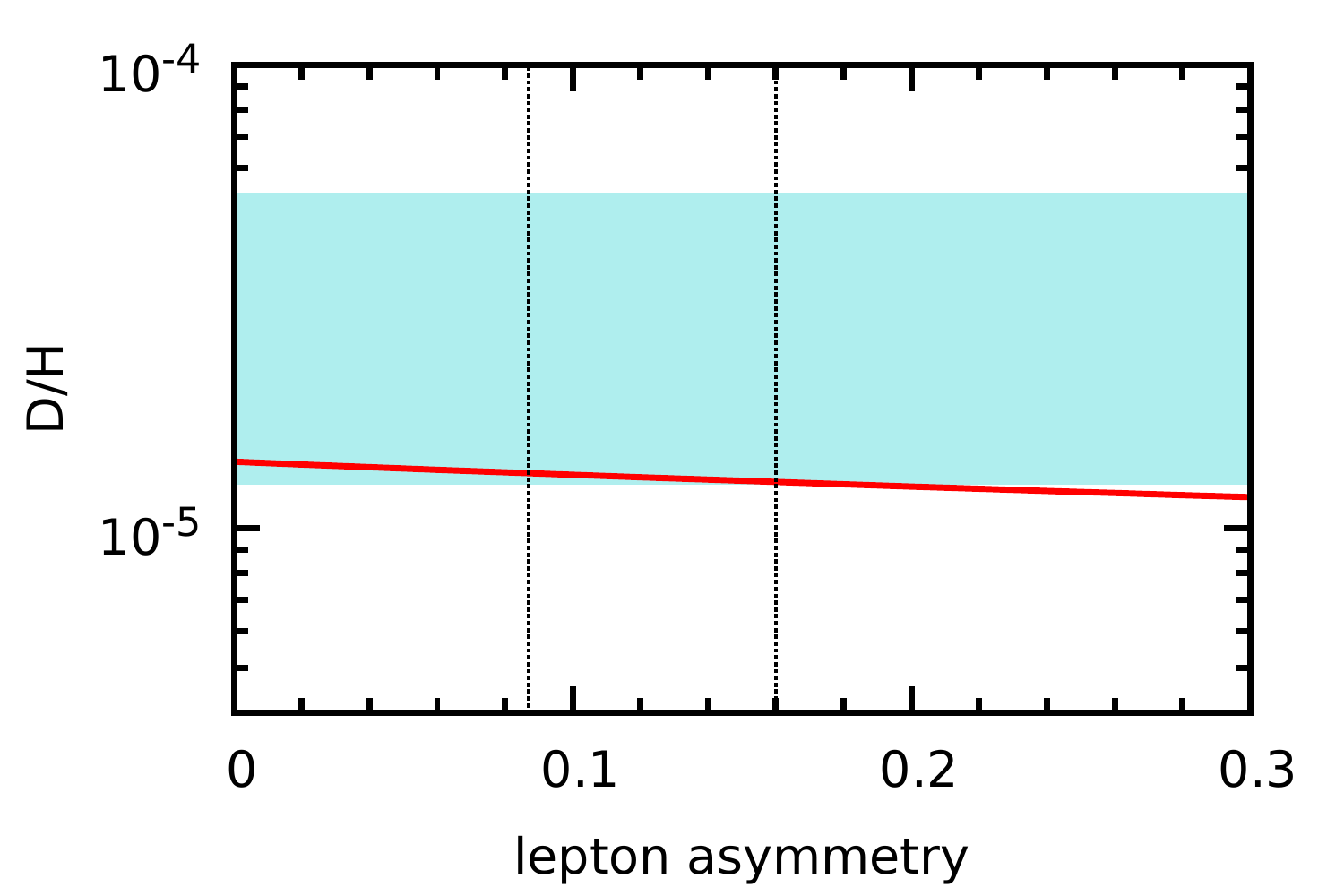}
\caption{{\it Left panel:} Helium abundance vs lepton asymmetry in our model is shown by solid curve. Gray band indicates observationally allowed range of helium abundance.  Range of $\chi$, required to match observations, is shown by vertical dotted lines.
{\it Right panel:} Deuterium abundance vs lepton asymmetry. Gray band indicates observationally allowed range of D/H. Vertical dotted lines are from the left panel.}
\label{fig:He-D}
\end{figure}

\begin{figure}
\includegraphics[width=0.5\textwidth]{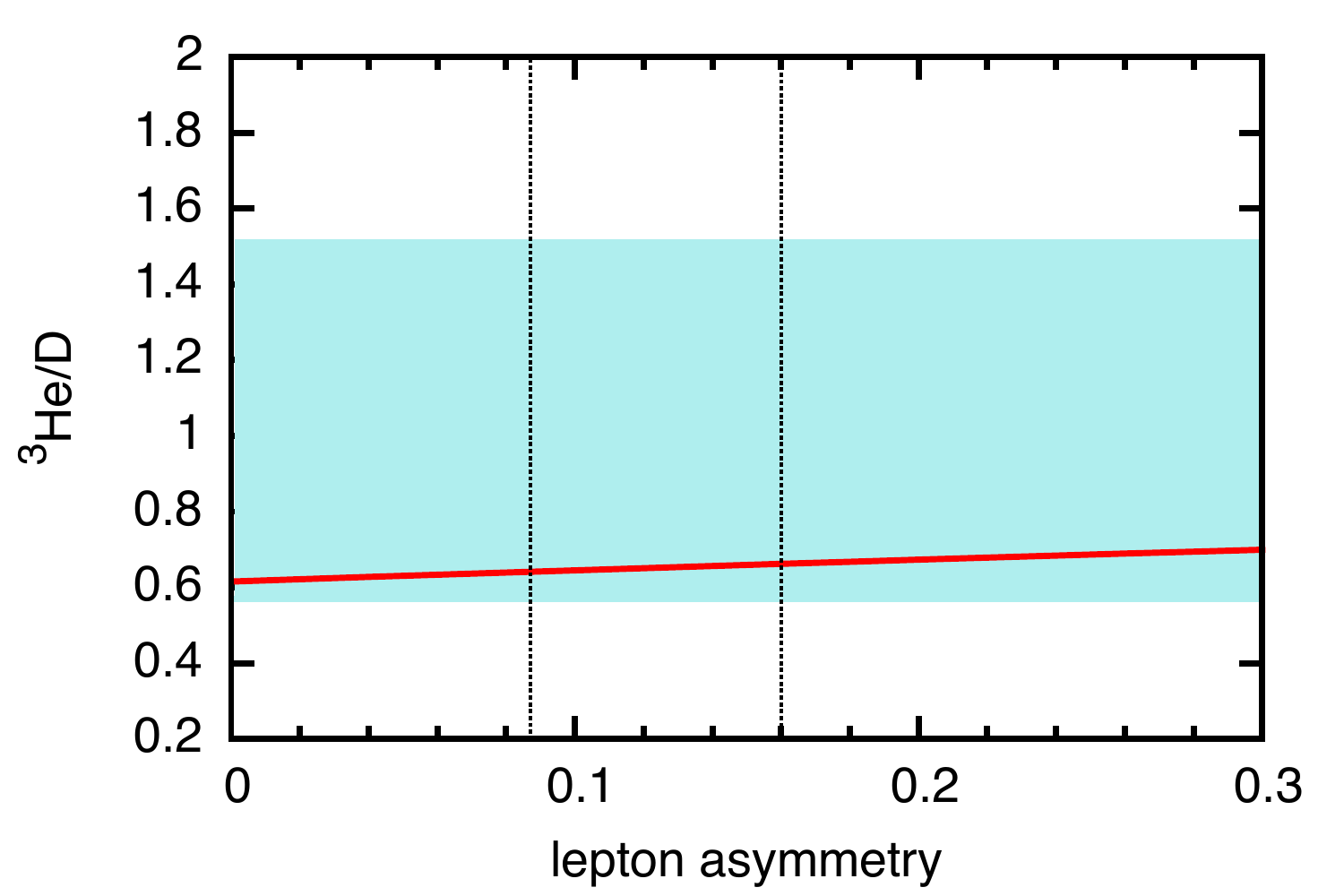}
\includegraphics[width=0.5\textwidth]{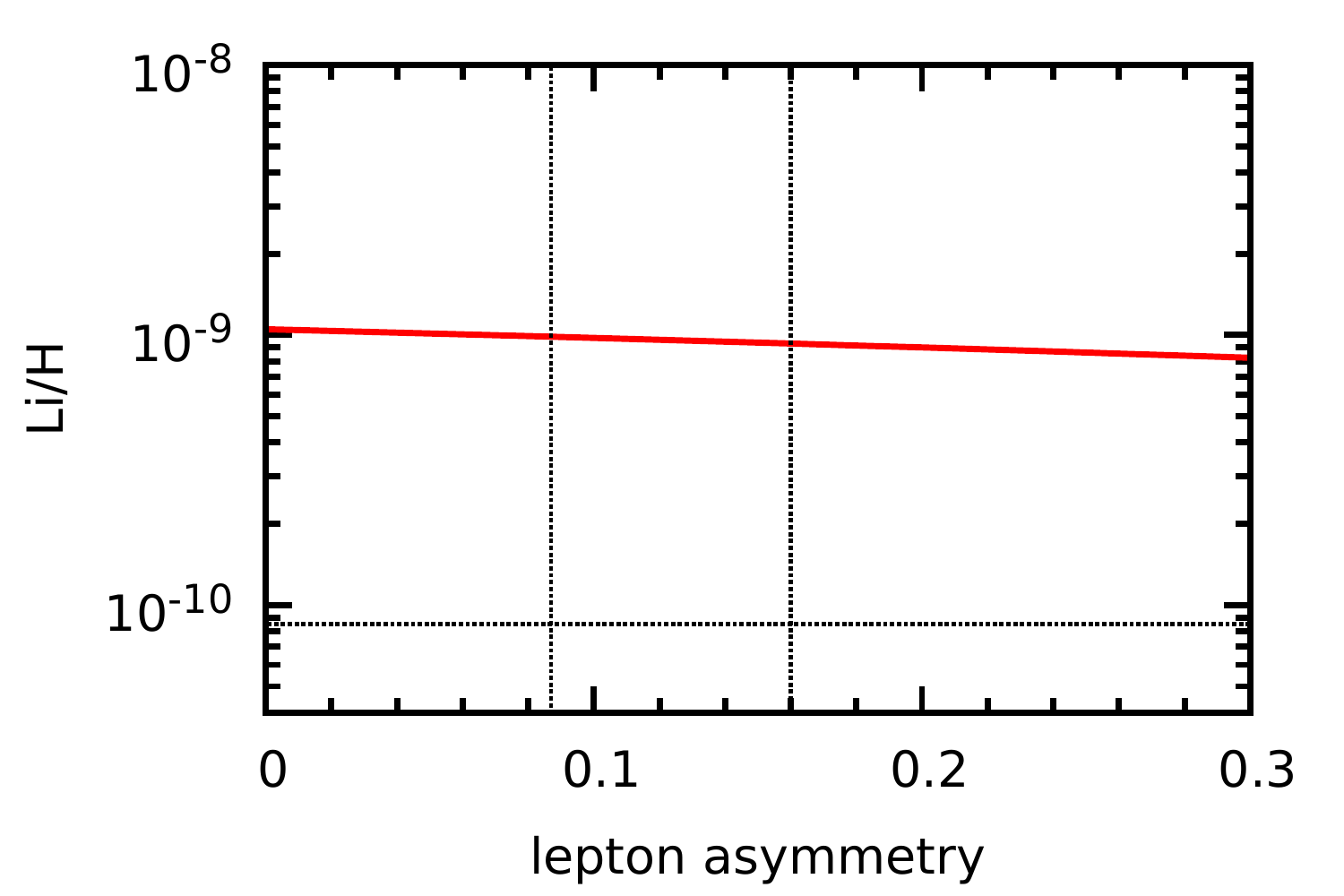}
\caption{{\it Left panel:}  $^3{\rm He/D}$ ratio vs lepton asymmetry. Gray band indicates observationally allowed range of this ratio. Vertical dotted lines are from Fig. 1.
{\it Right panel:} Li/H ratio vs lepton asymmetry. Observationally allowed range is above horizontal dotted line.}
\label{fig:H3-Li}
\end{figure}

We have calculated abundances numerically. To this end we have modified accordingly the publicly available numerical code of Kawano \cite{Kawano:1992ua}, so that all  effects mentioned above are automatically taken into account.

Following Refs. \cite{Jedamzik:2006xz,Arbey:2011nf} we use the conservative constraints on light elements abundances
\begin{eqnarray}
&&0.240 < Y_p <0.258, \label{eq:Y_p}\\ 
&&1.2\cdot 10^{-5} < D/H <5.3\cdot 10^{-5},\\ 
&&0.57 ~<~ ^3He/D ~<~ 1.52,\\
&&^7Li/H ~>~ 0.85\times10^{-10}.
\end{eqnarray}

Without lepton asymmetry the He abundance turns out to be in clear conflict with observations. In the standard model it is BBN which gives the strongest constraints on asymmetry between neutrinos and antineutrinos. Due to strong mixing between different neutrino species chemical potentials in all neutrino sectors are the same and in the standard model they are bounded from above  as $\chi \equiv \mu /T < 0.07$~\cite{mu-nu}. 

In the present case, on the contrary, we can introduce lepton asymmetry to bring Helium 
abundance in accord with observations. 

Calculated He abundance as a function of $\chi$ is shown in Fig.~\ref{fig:He-D}, left panel, for the case of charged scalar, $N=2$. Colored band represents 
the range given by inequality~ (\ref{eq:Y_p}), which in turn determines the required range of $\chi$ shown by vertical dotted lines.

Corresponding results for D/H, $^3{\rm He/D}$ and Li/H ratios are shown in Figs.~\ref{fig:He-D} and \ref{fig:H3-Li}. As we see, ther
calculated ratios are marginally  consistent with conservative constraints (3)-(6).
We should stress that $^7{\rm Li}$ abundance in our model is about 50\% above the standard BBN predictions. This  worsens the standard model problem with $^7{\rm Li}$ and we need even more stellar depletion that is typically invoked.


\subsection{Heavier dark matter, $m_X> m_e$}

Situation becomes very much different for heavier X-particles, especially if $m_X >m_e$, and was studied in detail in Ref.~\cite{Serpico:2004nm}. In this case the cosmological number density of $X$-particles would be determined by their annihilation into $e^+e^-$. 
Taking for the annihilation cross-section 
the approximate expression 
\be
v \sigma(X \bar X \rar e^+e^-) = \pi \alpha \alpha'  \tau / m_X^2,
\label{sigmaXXee}
\ee
where $\tau = \sqrt{1- 4m_e^2/m_X^2}$ and assuming for simplicity that $m_X\gg m_e$, we find:
\be
\rho_X^c \approx 2\cdot 10^{-2} \,\frac{{\rm keV}}{\rm{cm^3}} \, \left(\frac{ m_3^2}{\alpha_{11}}\right)  \, [1 + 0.05\,\ln (\alpha_{11}/m_3)] , 
\label{rho-X-c}
\ee
where $m_3 = m_X/{\rm MeV} $. 

If $m_X$ is sufficiently small, such that the equilibrium number density of $X$-particles is non-negligible at $n/p$-freezing, then
their presence would speed up the cosmological expansion and in this sense is equivalent to some dark radiation, though X-particles
at this stage were non-relativistic or at most semi-relativistic. On the other hand, additional $e^+e^-$-pairs from $X \bar X$-annihilation
would diminish relative contribution of neutrinos with respect to photons and this results in a decrease of $N_{eff}$. For example for
$m_X \simeq m_e$ the ratio of neutrino to photon temperature would be $(T_\nu/T_\gamma)^3 = 4/(11+2N)$ and if $N=2$ we
obtain $T_\nu/T_\gamma = 0.64$ already at BBN but not much later as it was in the case of light X-particles. So depending upon $m_X$ the overall effect of X-particles on BBN may be of either sign. E.g., as it was shown in Ref.~\cite{Serpico:2004nm}, the existence of millicharged particles in the mass range $m_X=4$ -- 10 MeV can even improve the overall agreement between the predicted and observed $^2$H and $^4$He abundances.


\section{Conclusion}

We have found that assuming non-zero lepton asymmetry one may avoid the BBN
upper bounds on the charge of possible light milli-charged particles. This in turn 
opens a way to modify the standard BBN predictions and to mimic possibly observed
dark  radiation. An interesting and testable feature of the model is that the values of the baryon and lepton asymmetries (i.e. the ratios $n_B/n_\gamma$ and $n_L/n_\gamma$) are different at BBN and recombination epochs.

The model discussed here was stimulated by our work on the role which X-particles might play in the generation of large scale magnetic fields after hydrogen recombination~\cite{magn-zb-ad-it}.

\acknowledgments 

A.D. and I.T. acknowledge the support of the Russian Federation Government Grant No. 11.G34.31.0047. The work of Z.B. and A.D. was supported in part by the MIUR biennal grant for the Research Projects of National Interest PRIN 2008
``Astroparticle Physics". The work of Z.B. was supported in part by the grant N14.U02.21.0913 of RF Ministery of Science and Education. The work of I.T. has been supported in part by the SCOPES program and  by the grant of the Russian Ministry of Education and Science No. 8412.

\end{document}